\documentclass[11pt,a4]{article} %% CFP says:  11pt at least, 10 pages.

\usepackage[T1]{fontenc}
\usepackage[utf8]{inputenc}
\usepackage{amssymb,amsmath}
\usepackage{times}
\usepackage{xspace}

\usepackage[left=3.4cm,top=4cm,bottom=4cm,right=3.4cm]{geometry}

\usepackage{authblk}

\usepackage{url}
\usepackage{listings,subfigure}

\newcommand{\x}{\xspace}
\newcommand{\coq}{\textsc{Coq}\x}

\newcommand{\setN}{\ensuremath{\mathbb{N}}\x}
\newcommand{\setR}{\ensuremath{\mathbb{R}}\x}

\renewcommand{\epsilon}{\varepsilon}
\usepackage{lstcoq}
\usepackage{tikz}

\tikzstyle{rond}=[circle=1pt,draw=black,line width=1pt]

\newcommand{\robot}[3]{\node (r) at (#1,#2) [draw,color=#3,fill,rond] {};}

\begin{document}

\title{A Certified Universal Gathering Algorithm \\ for Oblivious
  Mobile Robots} %
\author[1]{Pierre Courtieu}
\author[2]{Lionel Rieg}
\author[3,4]{Sébastien Tixeuil}
\author[5,1,6]{Xavier Urbain}
\affil[1]{\textsc{C\'edric} -- Conservatoire national des arts et
    m\'etiers, Paris, F-75141}
\affil[2]{Collège de France}
\affil[3]{Sorbonne Universités, UPMC Univ Paris 06, CNRS, LIP6 UMR 7606, 4 place Jussieu 75005 Paris}
\affil[4]{Institut Universitaire de France}
\affil[5]{\'Ecole Nat. Sup. d'Informatique pour l'Industrie
    et l'Entreprise (ENSIIE), \'Evry, F-91025}
\affil[6]{LRI, CNRS UMR 8623, Universit\'e Paris-Sud, Orsay, F-91405}
\renewcommand\Authands{ and }

\maketitle

\begin{abstract}
We present a new algorithm for the problem of universal gathering mobile
oblivious robots (that is, starting from any initial configuration
that is not bivalent, using any number of robots, the robots reach in a
finite number of steps the same position, not known beforehand) without relying on
a common chirality. 

We give very strong guaranties on the correctness of our algorithm by
\emph{proving formally} that it is correct, using the \coq proof
assistant. 

To our knowledge, this is the first certified positive (and
constructive) result in the context of oblivious mobile robots. It
demonstrates both the effectiveness of the approach to obtain new
algorithms that are truly generic, and its managability since the
amount of developped code remains human readable.  
\bigskip
\end{abstract}

\section{Introduction}
\label{sec:introduction}

\setcounter{page}{1}

Networks of mobile robots captured the attention of the distributed computing
community, as they promise new applications (rescue, exploration,
surveillance) in potentially dangerous (and harmful)
environments. Since its initial presentation~\cite{suzuki99siam}, this
computing model has grown in popularity and many refinements have been
proposed (see~\cite{flocchini12book} for a recent state of the art). From a
theoretical point of view, the interest lies in characterising the
exact conditions for solving a particular task.

In the model we consider, robots are anonymous (\emph{i.e.}, indistinguishable from each-other),
oblivious (\emph{i.e.}, no persistent memory of the past is available), and disoriented (\emph{i.e.}, they do not agree on a common coordinate
system). The robots operate in Look-Compute-Move cycles. In each cycle a robot ``Looks'' at its surroundings
and obtains (in its own coordinate system) a snapshot containing the
locations of all robots. Based on this visual information, the robot ``Computes'' a
destination location (still in its own coordinate system) and then ``Moves'' towards the computed
location. Since the robots are identical, they all follow the
same deterministic algorithm. The algorithm is oblivious if
the computed destination in each cycle depends only on the
snapshot obtained in the current cycle (and not on the past
history of execution). The snapshots obtained by the robots
are not consistently oriented in any manner (that is, the robots local
coordinate systems do not share a common direction nor a common chirality).

The execution model significantly impacts the solvability of collaborative tasks. Three different levels of synchronisation have
been considered. The strongest model~\cite{suzuki99siam} is
the fully synchronised (FSYNC) model where each phase
of each cycle is performed simultaneously by all robots.
On the other hand, the asynchronous model~\cite{flocchini12book}
(ASYNC) allows arbitrary delays between the Look, Compute and Move phases and the movement itself may take an
arbitrary amount of time. In this paper, we
consider the semi-synchronous (SSYNC) model~\cite{suzuki99siam}, which lies somewhere between the
two extreme models. In the SSYNC model, time is discretised
into rounds and in each round an arbitrary 
subset of the
robots are active. The robots that are active in a particular round perform exactly one atomic Look-Compute-Move cycle in
that round. It is assumed that the scheduler (seen as an adversary) is
fair in the sense that it guarantes that in any configuration, any robot is activated
within a finite number of steps.

\paragraph{Related Work.}

The gathering problem is one of the benchmarking tasks in mobile robot
networks, and has received a considerable amount of
attention (see~\cite{flocchini12book} and references herein). The gathering tasks consists in all robots
(considered as dimensionless points in a Euclidian space)
reaching
a single point, not known beforehand, in finite time. 
A foundational result~\cite{suzuki99siam,courtieu15ipl} shows that in the FSYNC or
SSYNC models, no oblivious deterministic
algorithm can solve gathering for two robots without additional
assumptions~\cite{izumi12siam}. 
This result can be extended~\cite{suzuki99siam,courtieu15ipl} to
the bivalent case, that is when an even number of robots is initially
evenly split in exactly two locations. On the other hand, it is
possible to solve gathering if $n>2$ robots start from initially
distinct positions, if robots are endowed with multiplicity detection:
that is, a robot is able to determine the number of robots that occupy
a given position. 
While probabilistic solutions~\cite{suzuki99siam,izumi13tpds} can cope with
arbitrary initial configuration (including bivalent ones), most of the
deterministic ones in the literature~\cite{flocchini12book} assume robots
always start from distinct locations (that is, the initial
configuration contains no multiplicity points). Some recent work was
devoted to relaxing this hypothesis in the deterministic case. 
Dieudonné and Petit~\cite{dieudonne12tcs} investigated the
problem of gathering from \emph{any} configuration (that is, the
initial configuration can contain arbitrary multiplicity points):
assuming that the number of robots is odd (so, no initial bivalent
configuration can exist), they provide a deterministic
algorithm for gathering starting from any configuration. Bouzid
\emph{et al.}~\cite{bouzid13icdcs} improved the result by also allowing an even number of
robots to start from configurations that contain multiplicity points
(albeit the initial bivalent configuration is still forbiden due to impossibility
results in this case). In that sense, the algorithm of Bouzid \emph{et
  al.}~\cite{bouzid13icdcs} is \emph{universal} in the sense that it works for all gatherable
configurations, including those with multiplicity points. Both
aforementioned results assume that robots and are endowed with multiplicity
detection and have a common chirality. A natural open question
emerging from those works is whether any of those assumptions can be
relaxed (not both of them can be relaxed at the same
time, as impossibility results exist in this case~\cite{prencipe07tcs}).

Another line of work that is related to our concern that of using
formal methods in the context of mobile robots~\cite{bonnet14wssr,devismes12sss,BMPTT13r,auger13sss,MPST14c,courtieu15ipl}. Model-checking
proved useful to find bugs in existing literature~\cite{BMPTT13r} and
assess formally published algorithms~\cite{devismes12sss,BMPTT13r}, in
a simpler setting where robots evolve in a \emph{discrete space} where
the number of possible positions is finite. 
Automatic program synthesis (for the problem of perpetual
exclusive exploration in a ring-shaped discrete space) is due to
Bonnet \emph{et al.}~\cite{bonnet14wssr}, and can be used to
obtain automatically algorihtms that are ``correct-by-design''. The approach was recently refined by Millet
\emph{et al.}~\cite{MPST14c} for the problem of gathering in a
discrete ring network. As all aforementioned approaches are designed
for a discrete setting where both the number of positions and the
number of robots are known, they cannot be used in the continuous space
where robots positions take values in a set that is not enumerable,
and they cannot permit to establish results that are valid for any
number of robots. 
Developed for the \coq proof
assistant,\footnote{\url{http://coq.inria.fr}} the Pactole framework enabled the use of high-order logic to certify
impossibility results~\cite{auger13sss} for the problem of
convergence: for any positive $\epsilon$, robots are required to reach
locations that are at most $\epsilon$ apart. Another classical
impossibility result that was certified using the Pactole framework is
the impossibility of gathering starting from a bivalent
configuration~\cite{courtieu15ipl}.
While the proof assistant approach seems a sensible path for
establishing certified results for mobile robots that evolve in a
continous space, to this paper there
exists no \emph{positive} certified result in this context. 
Expressing mobile robot
algorithms in a formal framework that permits certification poses a
double challenge: how to express the algorithm (that can make use of
complex geometric abstractions that must be properly defined within
the framework), and how to write the proof?

\subsubsection*{Our contribution}
Motivated by open problems on the gathering side and on the proof
assistant side, we investigate the possibility of \emph{universal} gathering mobile
oblivious robots (that is, starting from any initial configuration
that is not bivalent, using any number of robots) without relying on
chirality (unlike~\cite{dieudonne12tcs,bouzid13icdcs}). 

We present a new gathering algorithm for robots operating in a
continuous space that \emph{(i)} can start from any
configuration that is not bivalent, \emph{(ii)} does not put restriction on the
number of robots, \emph{(iii)} does not assume that robots share a
common chirality. 
We give very strong guaranties on the correctness of our algorithm by
\emph{proving formally} that it is correct, using the \coq proof
assistant. To this goal we use the formal model and libraries we
develop, and that has been previously sketched in \cite{auger13sss} and
\cite{courtieu15ipl}.

To our knowledge, this is the first certified positive (and
constructive) result in the context of oblivious mobile robots. It
demonstrates both the effectiveness of the approach to obtain new
algorithms that are truly generic, and its manageability since the
amount of developped code remains human readable. Our bottom-up
approach permits to lay sound theoretical foundations for future
developments in this domain.

\subsubsection*{Roadmap.} 
The sequel of the paper is organised as follows.
First, we recall the context of robot networks in
Section~\ref{sec:robots}. %, and we describe the problem of Gathering.
In Section~\ref{sec:unformal_robogram}, our algorithm is informally
presented, along with the key points of its correctness proof. We
present our formal \coq framework in Section~\ref{sec:formal},
together with the formalization of the key concepts identified in the
previous section. Section~\ref{sec:conclu} investigates further some
planned developments.  

The actual development
for \coq 8.5 is
available at\newline
\url{http://pactole.lri.fr/pub/certified_gathering1D.tgz}

\section{Robot Networks}\label{sec:robots}

We borrow most of the notions in this section
from~\cite{suzuki99siam,agmon2006fault,flocchini12book}.
The network consists in a set of $n$ mobile entities, called robots, arbitrarily located 
in the space.
Robots cannot communicate explicitely by sending messages to each others.
Instead, their communication is based on vision:
they observe the positions of other robots, and based on their observations, 
they compute destination points to which they move.

Robots are \emph{homogeneous} and \emph{anonymous}: 
they run the same algorithm (called \emph{robogram}),
they are completely 
indistinguishable by their appearance, and no 
identifier can be used in their computations.
They are also \emph{oblivious}, {i.e.} they cannot remember any previous observation, computation or movement
performed in any previous step.

For simplicity, we assume that robots are \emph{without volume},
\emph{i.e.}  they are modeled as points that cannot obstruct the
movement or vision of other robots.  Several robots can be located at
the same point, a \emph{tower} is a location inhabited by (one or)
several robots.  The multiplicity of a location $l$, that is the number
of robots at this location, is denoted by $|l|$.

Visibility is \emph{global}: the entire set of robots can always be
seen by any robot at any time.  Robots that are able to determine the
exact number of robots occupying a same position (i.e., the
multiplicity of a tower) enjoy \emph{strong} multiplicity detection; if
they can only know if a given position is inhabited or not, their
multiplicity detection is said to be \emph{weak}.

Each robot has its own local coordinate system and its own unit
measure. Robots do not share any origin, orientation, and more
generally any frame of reference, but it is assumed that every robot
is at the origin of its own frame of reference.

At a given time, robots and their positions define a
\emph{configuration}.
A configuration that consists of exactly two towers of same cardinalities is said to be \emph{bivalent}.

The degree of asynchrony in the robot swarm is characterised by
an abstract entity called the \emph{demon} (or
adversary).  Each time a robot is activated by the demon, it executes
a complete three-phases cycle: Look, Compute and Move.  During the
Look phase, using its visual sensors, the robot gets a snapshot of the
current configuration.  Then, based only on this observed
configuration, it computes a destination in the Compute phase using
its robogram, and moves towards it during the subsequent Move phase.
Movements of robots are \emph{rigid}, \emph{i.e.} the demon cannot
stop them before they reach the destination.

A \emph{run} (or execution) is an infinite sequence of rounds. 
During each round, the demon chooses a subset of robots and activates them to execute a cycle.
We assume the scheduling to be \emph{fair}, \emph{i.e.} each robot is
activated infinitely often in any infinite execution, 
and \emph{atomic} in the sense that robots that
are activated at the same round execute their actions synchronously and atomically.
An atomic demon is called fully-synchronous (FSYNC) if all robots are activated at each round, 
otherwise it is said to be semi-synchronous (SSYNC).

\section{Setting and Robogram}\label{sec:unformal_robogram}

We consider a set of $nG$ anonymous robots that are oblivious and
equipped with global strong multiplicity detection (that is, they are
able to count the number of robots that occupy any given position). The demon is
supposed to be fair, and the execution model is SSYNC. 

The space in which they move is the real line $\setR$. Robots do not
share any common direction of the line, nor any chirality. 

Any initial configuration is accepted as long as it is not
bivalent (including those with multiplicity points). 
Indeed, \cite{suzuki99siam} shows that
gathering is not solvable for two robots, and a formal certified proof that the
gathering problem cannot be solved if bivalent positions are tolerated
is available~\cite{courtieu15ipl}.

\subsection{Robogram}\label{sec:unformal_algo}

In this particular case of the considered space being $\setR$, even if there
is no common frame of reference, we have that, for any configuration,
the set of
inhabited locations that are the \emph{most external} is the same for
all robots.
Hence, those most external inhabited location define the same \emph{center of extrema} to all robots, as
well as the same set of (strictly) interior inhabited locations. 
Based on this remark, we can define the robogram embedded in each
robot as follows:

\begin{enumerate}
\item \label{algo:majo}If there is a unique location with highest
  multiplicity, the destination is that location,
\item\label{algo:three}Otherwise, if there are exactly three inhabited
    locations, the destination is the one in between,
\item\label{algo:center}Otherwise, if not already at one of the most external locations,
  the destination is the center of the most external ones.
\item Otherwise, the destination is the origin (do not
  move).
\end{enumerate}

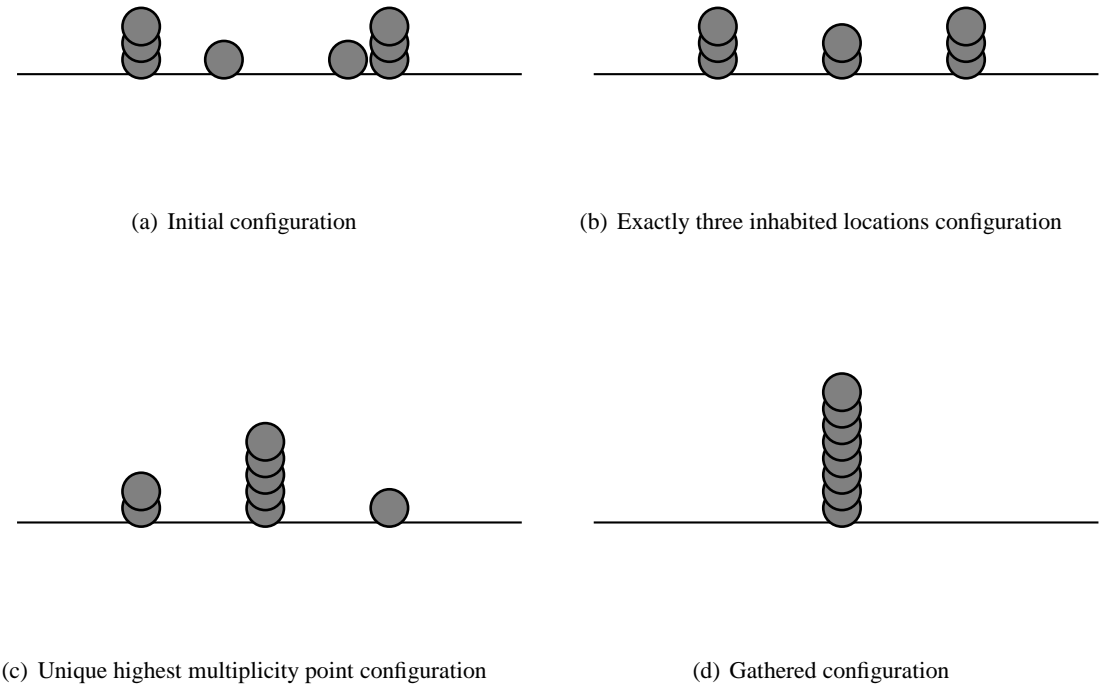
\begin{figure}[htb]
\subfigure[Initial configuration]{
    \begin{tikzpicture}[scale=1.1]
      \node(dummy) at (-2,-1.5) {};
      \node(dummy) at (+3,+3) {};
      \begin{scope}[>=stealth,inner sep=5pt, node distance=1cm,thick,line width=.8pt]
        \draw [color=black] (-1.5,-5pt) -- (4.6,-5pt);
        \robot{3}{0cm}{gray}
        \robot{3}{0.2cm}{gray}
        \robot{3}{0.4cm}{gray}
       \robot{1}{0cm}{gray}
       \robot{2.5}{0cm}{gray}

        \robot{0}{0cm}{gray}
        \robot{0}{.2cm}{gray}
        \robot{0}{.4cm}{gray}
   \end{scope}
   \end{tikzpicture}
}
\subfigure[Exactly three inhabited locations configuration]{
\begin{tikzpicture}[scale=1.1]
      \node(dummy) at (-2,-1.5) {};
      \node(dummy) at (+3,+3) {};
      \begin{scope}[>=stealth,inner sep=5pt, node distance=1cm,thick,line width=.8pt]
        \draw [color=black] (-1.5,-5pt) -- (4.6,-5pt);
        \robot{3}{0cm}{gray}
        \robot{3}{0.2cm}{gray}
        \robot{3}{0.4cm}{gray}
       \robot{1.5}{0cm}{gray}
       \robot{1.5}{0.2cm}{gray}

        \robot{0}{0cm}{gray}
        \robot{0}{.2cm}{gray}
        \robot{0}{.4cm}{gray}
   \end{scope}
   \end{tikzpicture}
}
\subfigure[Unique highest multiplicity point configuration]{
\begin{tikzpicture}[scale=1.1]
      \node(dummy) at (-2,-1.5) {};
      \node(dummy) at (+3,+3) {};
      \begin{scope}[>=stealth,inner sep=5pt, node distance=1cm,thick,line width=.8pt]
        \draw [color=black] (-1.5,-5pt) -- (4.6,-5pt);
        \robot{3}{0cm}{gray}
        
       \robot{1.5}{0cm}{gray}
       \robot{1.5}{0.2cm}{gray}
        \robot{1.5}{0.4cm}{gray}
        \robot{1.5}{0.6cm}{gray}
        \robot{1.5}{.8cm}{gray}
        \robot{0}{0cm}{gray}
        \robot{0}{.2cm}{gray}
   \end{scope}
   \end{tikzpicture}
}
\subfigure[Gathered configuration]{
\begin{tikzpicture}[scale=1.1]
      \node(dummy) at (-2,-1.5) {};
      \node(dummy) at (+3,+3) {};
      \begin{scope}[>=stealth,inner sep=5pt, node distance=1cm,thick,line width=.8pt]
        \draw [color=black] (-1.5,-5pt) -- (4.6,-5pt);
        \robot{1.5}{0cm}{gray}
        \robot{1.5}{0.2cm}{gray}
       \robot{1.5}{0.4cm}{gray}
       \robot{1.5}{0.6cm}{gray}

        \robot{1.5}{0.8cm}{gray}
        \robot{1.5}{1cm}{gray}
        \robot{1.5}{1.2cm}{gray}
        \robot{1.5}{1.4cm}{gray}
   \end{scope}
   \end{tikzpicture}
}
\caption{Example execution of our robogram.}
\label{fig:example}
\end{figure}

An example execution of our robogram is presented in
Figure~\ref{fig:example}. In the initial configuration (see
Figure~\ref{fig:example}.\emph{(a)}), only the third condition is
enabled. The inner robots move toward the middle of the extremal
robots. When there are three inhabited locations (see
Figure~\ref{fig:example}.\emph{(b)}), only the second condition is
enabled, and extremal robots move toward the inner inhabited
location. 
When a single highest multiplicity point is reached (see
Figure~\ref{fig:example}.\emph{(c)}), only the first condition is
enabled,and all robots move toward it. After all robots gather  (see
Figure~\ref{fig:example}.\emph{(d)}), only the fourth condition
apply, and the configuration is final. 
  
This description of the protocol is obviously informal, however we
present in Section~\ref{sec:robogram} its formal version,
that is, the \coq definition of our algorithm. 

\subsection{Key points to prove correctness}

Some properties are fundamental in our proof that the algorithm is a
solution to the problem of gathering. Namely, that robots move towards
the same location, that a legal configuration cannot evolve into a
forbidden (that is: bivalent) one, and finally that the configuation is
eventually reduced to a single inhabited location.

\subsubsection*{Robots that move go to the same location.}\label{key:same_dest}

Note that by robots ``that move'' we explicitely mean robots the
destination of which is not their original location, and \emph{not}
robots that are activated (some of which may not move). 
Robots enjoy global strong multiplicity detection, hence they all
detect if there is a unique tower with the highest multiplicity, thus
sharing the destination (Phase~\ref{algo:majo}).
If they do not find such a tower, they can all count how many
locations are inhabited. Should they detect that there are only three
of them (Phase~\ref{algo:three}) then, as previously remarked, sharing
the notion of tower in between, they also share the destination. Finally if there is
more than three inhabited locations none of which holding more robots
than the others (Phase~\ref{algo:center}), as most external towers are the
same for all robots, robots to move go the location defined as the
center of those external towers, that is the same destination again.

Further note that we actually just showed that all moving robots are
in the same phase of the robogram, and that the resulting destination does not depend on the
frame of reference of the robot. %

\subsubsection*{Bivalent positions are unreachable.}\label{key:not_biv}
We require that the initial configuration does not consist of exactly
two towers with the same multiplicity.  One of the key points ensuring
this algorithm's correctness is that there is no way to reach a position
that is bivalent from a position that is \emph{not}
bivalent.
Consider two configurations $C_0$ and $C_1$, $C_1$ being bivalent and resulting from
$C_0$ by some round. Let us denote by $|x|_0$ (resp. $|x|_1$) the
multiplicity of location $x$ in $C_0$ (resp. in $C_1$).
By definition, $C_1$ consists of two locations $l_1$ and $l_2$ such that $|l_1|_1 =
|l_2|_1=\frac{nG}{2}$. As all moving robots go to \emph{the same location}, we can assume
without loss of generality that robots moved to, say, $l_1$, adding to
its original multiplicity $|l_1|_0$ (which might have been $0$). % so that \emph{after} the round
                                % it equals $|l_2|$. 
Since the configuration is now bivalent, this means that %\emph{before} $R$
$l_2$ was inhabited in $C_0$  and such that $|l_2|_0 \geq \frac{nG}{2}$ (some robot in $l_2$ might have
moved to $l_1$). There cannot have been only one inhabited location $l$ distinct
from $l_2$ before the round because either $|l|_0=|l_2|_0=\frac{nG}{2}$ but we supposed
the configuration was not bivalent, or $|l|_0< \frac{nG}{2} < |l_2|_0$ but then by
Phase~\ref{algo:majo} robots would have moved to $l_2$ and not $l_1$.
Hence $C_0$ consisted of $l_2$ and several inhabited $l_i$ ($i\neq2$) amongst which the
robots not located in $l_2$ were distributed, but then none of the
$l_i$ could have held more than $\frac{nG}{2} -1$ robots, hence
Phase~\ref{algo:majo} should have applied and robots should have moved
to $l_2$, a contradiction.

\subsubsection*{Eventually no-one moves.}\label{key:termination}
The termination of the algorithm is ensured by the existence of a
measure decreasing at each round involving a moving
robot % 
for a well-founded
ordering. We then conclude using the assumption that the demon is
fair. %

The measure is defined as follows:  we map any
configuration $C_i$ to a $(p_i,m^{p_i}_i) \in \setN \times \setN$ such
that $p_i$ is the phase number of the moving robots, and:
\begin{itemize}
\item $m^1_i$ is the number of robots that are \emph{not at} the (unique) location of
  highest multiplicity,
\item $m^2_i$ is the number of robots that are \emph{not at} the
  inhabited location in between,
\item $m^3_i$ is the number of robots that are \emph{neither at} a
  most external location \emph{nor at} their center.
\end{itemize}

Let $>_\setN$ be the usual ordering on natural numbers, the relevant
ordering $\succ$ is defined as the lexicographic extension of
$>_\setN$ on pairs: 
\[
(p,m) \succ (p',m') \quad\textrm{iff}\quad  
\left\{
  \begin{array}{cl}
   \textrm{either~} & p >_\setN p',\\
   \textrm{or} & p =_\setN p' \textrm{ and } m >_\setN m'.
  \end{array}
\right.
\]
It is well-founded since $>_\setN$ is well-founded.
We show that for 
any round on a configuration $C_k$ resulting in
a \emph{different} configuration $C_{k+1}$, $(p_k,m^{p_k}_k) \succ
(p_{k+1},m^{p_{k+1}}_{k+1})$, hence proving that eventually there is
no more change in successive configurations.

\section{A Formal Model to Prove Robograms}\label{sec:formal}

To certify results and to guarantee the soundness of theorems, we use
\coq, a Curry-Howard-based interactive  proof assistant
enjoying a trustworthy kernel.
The (functional) language of \coq is a very expressive
$\lambda$-calculus: the \emph{Calculus of Inductive Constructions}
(CIC)~\cite{coquand90colog}. In this context, datatypes, objects, algorithms,
theorems and proofs can be expressed in a unified way, as terms.

The reader will find in~\cite{bertot04coqart} a very comprehensive overview
and good practices with reference to \coq.
Developing a proof in a proof assistant may nonetheless be tedious, or require
expertise from the user.
To make this task easier, we are actively developing (under the name
Pactole) a formal model, as
well as lemmas and theorems, to specify and certify results about
networks of autonomous mobile robots.
It is designed to be robust and flexible enough to express most of the
variety of assumptions in robots network, for example with reference
to the considered space: discrete or continuous, bounded or
unbounded\ldots

We do not expect the reader to be an expert in \coq but of course the
specification of a model for mobile robots in \coq requires some
knowledge of the proof assistant. %
We want to stress that the framework eases the developer's task. %
The notations and definitions we give hereafter should be simply read
as typed functional expressions. 

The formal model we rely on, as introduced in~\cite{auger13sss}, exceeds
our needs % 
as in particular it includes Byzantine robots, which are
irrelevant in the present work. 
The reader is invited to check that the actual code is almost
identical.

\subsection{The Formal Model}\label{sec:formal_model}
The Pactole model\footnote{Available at \url{http://pactole.lri.fr}} has been sketched
in~\cite{auger13sss,courtieu15ipl} {to which we refer for further
  details}; we recall here its main characteristics.

We use two important features of \coq: a formalism of
\emph{higher-order} to quantify over programs,
demons, etc., and the possibility to define \emph{inductive} and
\emph{coinductive} types~\cite{sangiorgi12book} to express
inductive and coinductive datatypes and properties.
Coinductive types are in particular of invaluable help to express
infinite behaviours, infinite datatypes and
properties on them, as we shall see with demons.

Robots are anonymous, however we need to identify some of them in the
proofs. %
Thus, we consider given a finite set of \emph{identifiers}, isomorphic
to a segment of $\setN$. We hereafter omit this set \lstinline!G! 
unless it is necessary to characterise the number of
robots. 
Robots are distributed in space, at places called \emph{locations}.
We call a \emph{configuration} a \emph{function} from a set of
identifiers to the space of locations. 
 The set of locations we
consider here is the real line \setR. %

Note that from that definition, there is information about identifiers
contained in configurations, in particular, equality between
configurations does \emph{not} simply boil down to the equality of the
multisets of inhabited locations.

Now if we are under the assumption that robots are anonymous and
indistinguishable, we have to make sure that those identifiers are not
used by the embedded algorithm. %

\paragraph*{Spectrum.}\label{sec:spect}
The computation of any robot's target location is based on the
perception they get from their environment, that is, in an SSYNC
execution scheme, from a configuration. 
The result of this observation may be more or less accurate, depending
on sensors' capabilities. %
A robot's perception of a configuration is called a \emph{spectrum}. %
To allow for different assumptions to be studied, we leave abstract
the type \emph{spectrum} (\lstinline!Spect.t!) and the notion of
spectrum of a position. %
Robograms will then output a location when given a spectrum (instead
of a configuration), thus guarantying that assumptions over sensors
are fulfilled. %
For instance, the spectrum for anonymous robots with \emph{weak} global multiplicity detection could be a set of inhabited locations, i.e., without any multiplicity information.
In a \emph{strong} global multiplicity setting, a multiset of inhabited locations is a suitable spectrum; that is what we use in this work.

In the following we will distinguish a \emph{demon} configuration
(resp.  spectrum), that is expressed in the global frame of reference,
from a
\emph{robot} configuration (resp. spectrum), that is expressed in the
robot's own frame of reference. At each step of the distributed
protocol (see definition of \lstinline!round!  below) the demon
configuration and spectrum are transformed (i.e., recentered, rotated
and scaled) into the considered robots ones before being given as
parameters to robograms. Depending on assumptions, the zoom and
rotation factors may be fixed for each robot or chosen by the demon at
each step. They may also be shared by all robots or not, etc.

\paragraph*{Robogram.}\label{sec:robogram}
Robograms may be naturally defined in a \emph{completely abstract
  manner}, without any concrete code, in our \coq model as follows.
They consist of an actual algorithm \lstinline!pgm! that takes a
spectrum as input and returns a location, and a compatibility property
\lstinline!pgm_compat!  stating that target locations are the same if
equivalent spectra are given (for some equivalence on spectra).
\begin{lstlisting}
Record robogram := {
  pgm :> Spect.t -> Location.t;
  pgm_compat : Proper (Spect.eq \Parrow Location.eq) pgm}.
\end{lstlisting}

Of course it is possible to instanciate the robogram by giving
a concrete definition of the program, and proving that the
compatibility property holds. In our case, the type of locations is
\lstinline!R.t! (from the \coq library on axiomatic reals) and the
program as described in Section~\ref{sec:unformal_algo} is:

\begin{lstlisting}
Definition robogram_pgm (s: Spect.t) : R.t :=
  match Spect.support (Smax s) with (* Locations of max multiplicity *)
    | nil => 0                       (* Only happens if no robot      *)
         if beq_nat (length (Spect.support s)) 3 then
           List.nth 1 (sort (Spect.support s)) 0  (* Phase 2: between*)
         else if is_extremal 0 s then 0           (* ... stay...     *)
              else extreme_center s               (* Phase 3: center *)
  end.
\end{lstlisting}
Note that this is almost exactly an ML code.

The resulting instanciated robogram is defined under the name \lstinline!gathering_robogram!.

\subsection{Formalising Key Points and the Main Theorem}

The key steps of our proof can be expressed as relatively
straightforward statements.
Theorem \lstinline!same_destination! states that two robots $id_1$ and
$id_2$ that are in the set of moving robots (i.e., the destination of
which is not their current location) compute the same destination
location (in the demons's frame of reference).
\begin{lstlisting}
Theorem same_destination : forall da config id1 id2,
  In id1 (moving gathering_robogram da config) 
  -> In id2 (moving gathering_robogram da config) 
  -> round gathering_robogram da config id1 =
      round gathering_robogram da config id2.
\end{lstlisting}

By case on the phases of the robogram, and on the structure of the
provided code. The formal proof is about 30 lines of \coq
long.

Theorem \lstinline!never_forbidden! says that for all demonic action
\emph{da} and configuration \emph{conf}, if \emph{conf} is not bivalent,
then the configuration resulting from \emph{conf} after the round
defined by \emph{da} and our robogram is not bivalent.
\begin{lstlisting}
Theorem never_forbidden : 
  forall da conf, \not forbidden conf 
  -> \not forbidden (round gathering_robogram da conf).
\end{lstlisting}
Proof is done by a case analysis on the set of towers of maximum
height at the beginning. %
If there is none, this is absurd; %
if there is exactly one, the resulting configuration will have the
same highest tower, a legal configuration. %
Now if there are at least two highest towers, then if the resulting
configuration is bivalent, at least one robot has moved (otherwise the
original configuration would be bivalent, to the contrary of what is
assumed), and all robots that move go to the same of the resulting two
towers. The rest is arithmetics, as described on
page~\pageref{key:not_biv}. The proof of this key point is less than 100
lines of \coq script.

It remains to state that for all demonic action \emph{da} and configuration
\emph{conf}, if \emph{conf} is not bivalent, and if there is at least
one robot moving, then the configuration resulting from the round
defined by \emph{da} and our robogram on \emph{conf} is smaller than
\emph{conf}. The ordering relation on configurations, called
\lstinline!lt_conf!, being the one described in
section~\ref{key:termination}.  This is directly translated into the
following theorem. 

\begin{lstlisting}
Theorem round_lt_conf : forall da conf,
  \not forbidden conf -> moving gathering_robogram da conf <> nil 
  -> lt_conf (round robogram da conf) conf.
\end{lstlisting}

A general description on how to characterise a solution to the problem
of gathering has been given in \cite{courtieu15ipl}. We specialise
this definition here to take into account that an initial
configuration is not bivalent. This is straightforward: any robogram
$r$ is a solution w.r.t. a demon $d$ if for
all configuration \emph{conf} that is not bivalent, there is a point
$pt$ to which all robots will eventually gather (and stay) in the execution
defined by $r$ and $d$, and starting from \emph{conf}.
\begin{lstlisting}
Definition solGathering (r : robogram) (d : demon) :=
  forall conf, \not forbidden conf -> exists pt : R, WillGather pt (execute r d conf).
\end{lstlisting}

The theorem stating the correctness of our robogram is then simply:
for all demon $d$ that is fair, 
\lstinline!gathering_robogram! is a solution with reference to $d$.
\begin{lstlisting}
Theorem Gathering_in_R :
  forall d, Fair d -> solGathering gathering_robogram d.
\end{lstlisting}

The proof is led by well-founded induction on the \lstinline!lt_conf!
relation. If all robots are gathered, then it is done. If not, by fairness some robots will have to move, thus a robot will be amongst the first to move. (Formally, this is an induction using fairness.)
We conclude by using the induction hypothesis (of our well-founded induction) as this round decreases the measure on configurations (theorem \lstinline!round_lt_conf!).
This proof of the main theorem is interestingly small as it is only about 20 lines of \coq.

The whole file dedicated to specification and certification of our algorithm
(\lstinline!RDVinR.v!) is
about 2300~lines long.
It includes 460~lines of definitions, specification and intermediate lemmas, and approximately 1460~lines of actual proof.

\section{Perspectives}\label{sec:conclu}

We proposed a new algorithm to gather anonymous and oblivious robots
on a continuous unbounded space: the real line $\setR$, without
relying on a shared orientation or chirality, and allowing
for any initial configuration that is not bivalent. This protocol is
certified correct for any positive number of robots (more than $2$) using our actively
developed \coq framework for networks of mobile robots, which is
publicly available to the research community.

A next step would be to add more dimensions to the considered
Euclidian space, first by considering gathering in
$\setR^2$. As the framework is highly parametric, specifying another
space in which robots move is not a dramatic change: the type of
locations is a parameter, it is left abstract throughout the majority
of the formalism, in which a concrete instance is not
needed.

Another interesting evolution would be to take into account the more
general ASYNC model, that is when Look-Compute-Move cycles and phases
are not atomic anymore. %
Describing behaviours that are ASYNC in \coq may nonetheless
add to the intricacy of formal proofs, and relevant libraries to ease
the task of the developer will have to be provided accordingly. 

\newpage

\bibliographystyle{plain}

\end{document}